\documentclass[preprintnumbers,amsmath,amssymbm,prd]{revtex4}
\usepackage{epsfig}
\usepackage{graphicx}
\usepackage{amssymb}

\begin{document}
\title{Extremal Kerr-Newman black holes with extremely short
charged scalar hair}
\author{Shahar Hod}
\affiliation{The Ruppin Academic Center, Emeq Hefer 40250, Israel}
\affiliation{ } \affiliation{The Hadassah Institute, Jerusalem
91010, Israel}
\date{\today}

\begin{abstract}
The recently proved `no short hair' theorem asserts that, if a
spherically-symmetric static black hole has hair, then this hair
(the external fields) must extend beyond the null circular geodesic
(the``photonsphere") of the corresponding black-hole spacetime:
$r_{\text{field}}>r_{\text{null}}$. In this paper we provide
compelling evidence that the bound can be {\it violated} by {\it
non}-spherically symmetric hairy black-hole configurations. To that
end, we analytically explore the physical properties of cloudy
Kerr-Newman black-hole spacetimes -- charged rotating black holes
which support linearized stationary charged scalar configurations in
their exterior regions. In particular, for given parameters
$\{M,Q,J\}$ of the central black hole, we find the dimensionless
ratio $q/\mu$ of the field parameters which {\it minimizes} the
effective lengths (radii) of the exterior stationary charged scalar
configurations (here $\{M,Q,J\}$ are respectively the mass, charge,
and angular momentum of the black hole, and $\{\mu,q\}$ are
respectively the mass and charge coupling constant of the linearized
scalar field). This allows us to prove explicitly that
(non-spherically symmetric non-static) composed
Kerr-Newman-charged-scalar-field configurations can violate the
no-short-hair lower bound. In particular, it is shown that extremely
compact stationary charged scalar `clouds', made of linearized
charged massive scalar fields with the property $r_{\text{field}}\to
r_{\text{H}}$, can be supported in the exterior spacetime regions of
extremal Kerr-Newman black holes (here $r_{\text{field}}$ is the
peak location of the stationary scalar configuration and
$r_{\text{H}}$ is the black-hole horizon radius). Furthermore, we
prove that these remarkably compact stationary field configurations
exist in the {\it entire} range $s\equiv J/M^2\in (0,1)$ of the
dimensionless black-hole angular momentum. In particular, in the
large-mass limit they are characterized by the simple dimensionless
ratio ${{q}/{\mu}}={{(1-2s^2)}/{(1-s^2)}}$.
\end{abstract}
\bigskip
\maketitle

\section {Introduction.}

Wheeler's famous conjecture that ``black holes have no hair"
\cite{Whee,Car} predicts a simple and universal fate for all
dynamical black-hole spacetimes \cite{Noteflat}: the matter fields
outside the horizon are expected to be swallowed by the black hole
or to be scattered away to infinity, thus leaving behind a
stationary ``bald" Kerr-Newman black hole \cite{Chan,Kerr,Newman}.
The no-hair conjecture therefore suggests that, within the framework
of classical general relativity, black holes are fundamental objects
which possess only three conserved physical parameters: mass $M$,
charge $Q$, and angular momentum $J$.

The no-hair conjecture predicts, in particular, that asymptotically
flat black holes cannot support static matter configurations in
their exterior regions. Early studies of the coupled Einstein-matter
equations have indeed ruled out the existence of regular black-hole
solutions with static scalar hair \cite{Chas}, static spinor hair
\cite{Hart}, and static massive vector hair \cite{BekVec}. These
early no-hair theorems have therefore supported the simple physical
picture suggested by the no-hair conjecture \cite{Whee,Car}.

However, the somewhat surprising discovery of regular black-hole
solutions \cite{Notereg} to the Einstein-Yang-Mills equations
\cite{BizCol} has revealed that coupled Einstein-matter systems may
exhibit a more complex behavior. The numerical discovery of these
`colored' black holes \cite{BizCol}, which provided the first
genuine counterexample to the no-hair conjecture, has motivated many
researches to search for other types of non trivial hairy black-hole
configurations. In fact, it is by now well established that black
holes can support various types of non-linear matter fields (that
is, matter fields with self-interaction terms) in their exterior
regions
\cite{BizCol,Lavr,BizCham,Green,Stra,BiWa,EYMH,Volkov,BiCh,Lav1,Lav2,Bizw}.

The hairy black-hole solutions discovered numerically in
\cite{BizCol,Lavr,BizCham,Green,Stra,BiWa,EYMH,Volkov,BiCh,Lav1,Lav2,Bizw}
provide compelling evidence that the no-hair conjecture, in its
original formulation \cite{Whee,Car}, may be violated
\cite{Notevio}. Accepting the fact that hairy black-hole
configurations do exist in general relativity, it is natural to
consider the following interesting question: What are the {\it
generic} features of hairy black-hole spacetimes?

This question was partially addressed in \cite{Hod11}, where a `{\it
no short hair}' theorem for spherically-symmetric static black holes
was proved. This no-short-hair theorem asserts that, if a
spherically-symmetric static black hole has hair, then this hair
(i.e., the external matter fields) must extend beyond the null
circular geodesic (the``photonsphere") of the corresponding
black-hole spacetime:
\begin{equation}\label{Eq1}
r_{\text{field}}>r_{\text{null}}\  .
\end{equation}
It is worth noting that, within the static sector of
spherically-symmetric hairy black-hole spacetimes, the no-short-hair
lower bound (\ref{Eq1}) is universal in the sense that it is
independent of the parameters of the exterior matter fields
\cite{Hod11,Hodaa,Notenun}. One may therefore regard the
no-short-hair relation (\ref{Eq1}) as a more modest alternative (at
least within the spherically-symmetric sector of hairy black-hole
configurations) to the original \cite{Whee,Car} no hair conjecture.

It is important to stress the fact that the formal proof provided in
\cite{Hod11} for the existence of the no-short-hair property
(\ref{Eq1}) is restricted to the relatively simple case of
spherically-symmetric static hairy black-hole spacetimes. The main
goal of the present paper is to challenge the validity of this
no-short-hair relation (\ref{Eq1}) beyond the regime of
spherically-symmetric static black holes. To that end, we shall
study analytically the physical properties of {\it non}-spherically
symmetric {\it non}-static Kerr-Newman black holes linearly coupled
to stationary (rather than static) charged massive scalar fields.

\section{Composed black-hole-scalar-field configurations.}

It is important to emphasize that while existing no-hair theorems
\cite{Chas} rigorously rule out the existence of static regular
hairy black-hole-scalar-field configurations, they do {\it not} rule
out the existence of non-static scalar field configurations in the
exterior regions of black-hole spacetimes.

In fact, recent analytical \cite{Hodrc} and numerical \cite{HerR}
explorations of the Einstein-scalar and Einstein-Maxwell-scalar
equations have revealed that stationary configurations of massive
scalar fields (with or without electric charge) can be supported in
the exterior spacetime regions of non-spherically symmetric {\it
rotating} black holes.

These non-static spatially regular black-hole-scalar-field
configurations \cite{Hodrc,HerR} owe their existence to the well
established phenomenon of superradiant scattering
\cite{Zel,PressTeu1,Bekad} of bosonic fields in black-hole
spacetimes \cite{Noterc}. In particular, these exterior stationary
field configurations have azimuthal frequencies
$\omega_{\text{field}}$ which coincide with the critical (threshold)
frequency $\omega_{\text{c}}$ for superradiant scattering in the
charged rotating black-hole spacetime \cite{Noteunits}:
\begin{equation}\label{Eq2}
\omega_{\text{field}}=\omega_{\text{c}}\equiv
m\Omega_{\text{H}}+q\Phi_{\text{H}}\ ,
\end{equation}
where \cite{Chan,Kerr,Newman,Notepar}
\begin{equation}\label{Eq3}
\Omega_{\text{H}}={{a}\over{r^2_++a^2}}\ \ \ {\text{and}} \ \ \
\Phi_{\text{H}}={{Qr_+}\over{r^2_++a^2}}
\end{equation}
are the angular velocity and the electric potential of the
Kerr-Newman black hole, and $\{m,q\}$ are respectively the azimuthal
harmonic index and charge coupling constant of the scalar field
\cite{Notesmp}.

The resonance condition (\ref{Eq2}) guarantees that the orbiting
scalar field is not absorbed by the black hole
\cite{Hodrc,HerR,Dolh}. In addition, for an asymptotically flat
black hole to be able to support {\it stationary} (that is,
non-decaying) field configurations in its exterior region, the
bounded fields must be prevented from radiating their energy to
infinity. For massive fields the required confinement mechanism is
naturally provided by the mutual gravitational attraction between
the central black hole and the orbiting massive bosonic
configuration. In particular, bound-state (that is, asymptotically
decaying) eigenfunctions of a scalar field of mass $\mu$
\cite{Notedim} are characterized by low frequency modes in the
regime [see Eq. (\ref{Eq17}) below]
\begin{equation}\label{Eq4}
\omega^2<\mu^2\  .
\end{equation}

As discussed above, the no-short-hair property (\ref{Eq1}) was
rigorously proved in \cite{Hod11} for the particular case of
spherically-symmetric static hairy black-hole configurations. The
main goal of the present paper is to challenge the validity of this
relation beyond the regime of spherically-symmetric static black
holes. To that end, we shall here study analytically the physical
properties of the (non-static non-spherically symmetric) composed
Kerr-Newman-charged-massive-scalar-field configurations
\cite{Hodrc,HerR} in the large-mass regime \cite{Notemm}
\begin{equation}\label{Eq5}
M\mu\gg1\  .
\end{equation}

Before proceeding, it is worth emphasizing that the composed
black-hole-scalar-field configurations that we shall analyze here
are not genuine hairy black-hole configurations. In particular, the
exterior charged massive scalar fields will be treated at the {\it
linear} level. We shall therefore use the term `clouds' to describe
these linearized exterior scalar configurations (the term `hair'
usually describes non-linear exterior fields).
As we shall show below, the main advantage of the present approach
lies in the fact that the composed
black-hole-linearized-scalar-field system is amenable to an {\it
analytical} treatment \cite{Notefn}.

\section{Description of the system.}

We study a physical system which is composed of a charged massive
scalar field $\Psi$ linearly coupled to an extremal charged rotating
Kerr-Newman black hole of mass $M$, electric charge $Q$, and
angular-momentum per unit mass $a$ \cite{Notepst}. Using the
Boyer-Lindquist coordinate system, the spacetime metric is described
by the line element \cite{Chan,Kerr,Newman}
\begin{eqnarray}\label{Eq6}
ds^2=-{{\Delta}\over{\rho^2}}(dt-a\sin^2\theta
d\phi)^2+{{\rho^2}\over{\Delta}}dr^2+\rho^2
d\theta^2+{{\sin^2\theta}\over{\rho^2}}\big[a
dt-(r^2+a^2)d\phi\big]^2\  ,
\end{eqnarray}
where $\Delta\equiv r^2-2Mr+a^2+Q^2$ and $\rho\equiv
r^2+a^2\cos^2\theta$. Extremal Kerr-Newman black holes are
characterized by the relation
\begin{equation}\label{Eq7}
r_{\text{H}}=M=\sqrt{a^2+Q^2}\  ,
\end{equation}
where $r_{\text{H}}$ is the radius of the degenerate black-hole
horizon.

The dynamics of a scalar field $\Psi$ of mass $\mu$ and charge
coupling constant $q$ \cite{Notedim} in the Kerr-Newman black-hole
spacetime is governed by the Klein-Gordon wave equation
\begin{equation}\label{Eq8}
[(\nabla^\nu-iqA^\nu)(\nabla_{\nu}-iqA_{\nu}) -\mu^2]\Psi=0\  ,
\end{equation}
where $A_{\nu}$ is the electromagnetic potential of the charged
black-hole spacetime. Substituting the field decomposition
\cite{Noteanz}
\begin{equation}\label{Eq9}
\Psi(t,r,\theta,\phi)=\sum_{l,m}e^{im\phi}{S_{lm}}(\theta;m,a\sqrt{\mu^2-\omega^2_{\text{c}}})
{R_{lm}}(r;M,Q,a,\mu,q,\omega_{\text{c}})e^{-i\omega_{\text{c}} t}\
\end{equation}
into the Klein-Gordon wave equation (\ref{Eq8}) and using the
Kerr-Newman metric components (\ref{Eq6}), one obtains two coupled
ordinary differential equations [see Eqs. (\ref{Eq10}) and
(\ref{Eq15}) below] for the radial and angular components,
$R_{lm}(r)$ and $S_{lm}(\theta)$, of the scalar eigenfunction
$\Psi$.

The angular equation for the familiar spheroidal harmonic
eigenfunctions ${S_{lm}}(\theta;m,s\epsilon)$ is given by
\cite{Stro,Heun,Fiz1,Teuk,Abram,Hodasy}
\begin{eqnarray}\label{Eq10}
{1\over {\sin\theta}}{{d}\over{\theta}}\Big(\sin\theta {{d
S_{lm}}\over{d\theta}}\Big)
+\Big[K_{lm}+({s}\epsilon)^2\sin^2\theta-{{m^2}\over{\sin^2\theta}}\Big]S_{lm}=0\
.
\end{eqnarray}
Here
\begin{equation}\label{Eq11}
s\equiv {{a}\over{M}}
\end{equation}
is the dimensionless spin (angular-momentum) of the Kerr-Newman
black hole, and the dimensionless physical quantity
\begin{equation}\label{Eq12}
\epsilon\equiv M\sqrt{\mu^2-\omega^2_{\text{c}}}\
\end{equation}
measures the deviation of the scalar field mass $\mu$ from the
resonant oscillation frequency $\omega_{\text{c}}$ [see Eq.
(\ref{Eq2})] of the field mode.

The discrete set of angular eigenvalues $\{K_{lm}(s\epsilon)\}$ is
determined from the regularity requirement of the corresponding
angular eigenfunctions (spheroidal harmonics)
$S_{lm}(\theta;s\epsilon)$ at the two poles $\theta=0$ and
$\theta=\pi$. We shall henceforth consider equatorial scalar modes
in the eikonal regime
\begin{equation}\label{Eq13}
l=m\gg1\  ,
\end{equation}
in which case the angular eigenvalues are given by
\cite{Yang,Hodpp,Noteangu}
\begin{equation}\label{Eq14}
K_{mm}(s\epsilon)=m^2+\sqrt{m^2+(s\epsilon)^2}-({s}\epsilon)^2+O(1)\
.
\end{equation}

The radial eigenfunctions ${R_{lm}}$ are determined by the Teukolsky
radial equation \cite{Teuk,Stro}
\begin{equation}\label{Eq15}
{{d}
\over{dr}}\Big(\Delta{{dR_{lm}}\over{dr}}\Big)+\Big[{{H^2}\over{\Delta}}
+2ma\omega_{\text{c}}-\mu^2(r^2+a^2)-K_{lm}\Big]R_{lm}=0\ ,
\end{equation}
where
\begin{equation}\label{Eq16}
H\equiv (r^2+a^2)\omega_{\text{c}}-am-qQr\  .
\end{equation}
Note that the angular eigenvalues $\{{K_{lm}}(s\epsilon)\}$
\cite{Notebr} couple the radial equation (\ref{Eq15}) to the angular
equation (\ref{Eq10}).

The bound-state resonances of the composed
black-hole-charged-massive-scalar-field system are characterized by
exponentially decaying (bounded) radial eigenfunctions at asymptotic
infinity \cite{Hodrc,HerR,Ins2}:
\begin{equation}\label{Eq17}
R(r\to\infty)\sim e^{-\epsilon r/r_{\text{H}}}\
\end{equation}
with $\epsilon^2>0$ \cite{Noteepp}. In addition, regular scalar
configurations are characterized by finite radial eigenfunctions. In
particular,
\begin{equation}\label{Eq18}
R(r=r_{\text{H}})<\infty\  .
\end{equation}

The physically motivated boundary conditions (\ref{Eq17}) and
(\ref{Eq18}), together with the resonance condition (\ref{Eq2}),
single out a discrete set of eigen field-masses \cite{Notenr}
$\{\mu_n(M,Q,a,l,m,q)\}^{n=\infty}_{n=0}$ (along with the associated
radial eigenfunctions) which characterize the stationary bound-state
resonances of the composed Kerr-Newman-charged-massive-scalar-field
system.

\section{The effective binding potential of the black-hole spacetime}

Before solving the radial Teukolsky equation (\ref{Eq15}), it proves
useful to analyze the spatial behavior of the effective radial
potential which binds the charged massive scalar field to the
charged rotating Kerr-Newman black hole. To that end, it is
convenient to define the new radial function
\begin{equation}\label{Eq19}
\psi=xR\  ,
\end{equation}
in terms of which the radial Teukolsky equation (\ref{Eq15}) can be
written in the form of a Schr\"odinger-like wave equation
\begin{equation}\label{Eq20}
{{d^2\psi}\over{dx^2}}-V\psi=0\  ,
\end{equation}
where
\begin{equation}\label{Eq21}
x\equiv {{r-M}\over {M}}\
\end{equation}
is a dimensionless radial coordinate. The effective radial potential
in (\ref{Eq20}) is given by
\begin{equation}\label{Eq22}
V=V(x;M,Q,a,l,m,\mu,q)=\epsilon^2-{{2\kappa\epsilon}\over{x}}+{{\beta^2-{1\over4}}\over{x^2}}\
,
\end{equation}
where
\begin{equation}\label{Eq23}
\kappa\equiv
{{M\omega_{\text{c}}(2M\omega_{\text{c}}-qQ)-(M\mu)^2}\over{\epsilon}}
\end{equation}
and
\begin{equation}\label{Eq24}
\beta^2\equiv{K+{1\over
4}-2Mm{s}\omega_{\text{c}}-(2M\omega_{\text{c}}-qQ)^2+(M\mu)^2(1+{s}^2)}\
.
\end{equation}

The boundary condition (\ref{Eq18}) together with the relation
(\ref{Eq19}) dictate
\begin{equation}\label{Eq25}
\psi(x=0)=0
\end{equation}
at the black-hole horizon. Thus, the effective radial potential
(\ref{Eq22}) of the Schr\"odinger-like wave equation (\ref{Eq20})
must be infinitely repulsive at $x=0$ \cite{Note1o2}:
\begin{equation}\label{Eq26}
V(x\to 0)\to +\infty\  .
\end{equation}
This implies that the stationary bound-state resonances of the
charged massive scalar fields in the extremal Kerr-Newman black-hole
spacetime are characterized by the inequality [see Eqs. (\ref{Eq22})
and (\ref{Eq26})] \cite{Notebeta,Note1o2}
\begin{equation}\label{Eq27}
\beta\geq{1\over 2}\  .
\end{equation}

In the case (\ref{Eq27}), and provided $\kappa>0$, the effective
radial potential (\ref{Eq22}) takes the form of a {\it trapping}
potential well which can support the stationary bound-state
resonances of the composed black-hole-charged-massive-scalar-field
system. In particular, in this case the binding potential
(\ref{Eq22}) has one minimum
which is located at
\begin{equation}\label{Eq28}
x_{\text{min}}={{\beta^2-{1\over 4}}\over{\kappa\epsilon}}\  .
\end{equation}

\section{The stationary bound-state resonances of the composed
Kerr-Newman-charged-massive-scalar-field system.}

In the present section we shall derive a (remarkably simple)
analytical formula for the discrete spectrum of field masses,
$\{\mu_n(M,Q,a,l,m,q)\}^{n=\infty}_{n=0}$, which characterize the
bound-state resonances (the stationary charged scalar clouds) of the
composed Kerr-Newman-charged-massive-scalar-field system

The solution of the radial equation (\ref{Eq15}) can be expressed in
the simple form \cite{Abram,Hodrc}:
\begin{equation}\label{Eq29}
R(x)=C_1\times x^{-{1\over 2}+\beta}e^{-\epsilon x}M({1\over
2}+\beta-\kappa,1+2\beta,2\epsilon x)+C_2\times(\beta\to -\beta)\  ,
\end{equation}
where $M(a,b,z)$ is the confluent hypergeometric function
\cite{Abram} and $\{C_1,C_2\}$ are normalization constants. The
notation $(\beta\to -\beta)$ in (\ref{Eq29}) means ``replace $\beta$
by $-\beta$ in the preceding term."

In order to obtain the resonance equation which characterizes the
bound-state resonances of the composed
Kerr-Newman-charged-massive-scalar-field system, we shall now
examine the asymptotic spatial behaviors of the radial eigenfunction
$R(x)$ in the limits $x\to 0$ and $x\to\infty$:
\newline
(1) The behavior of the radial eigenfunction (\ref{Eq29}) in the
near-horizon $x\ll 1$ region is given by \cite{Abram}
\begin{equation}\label{Eq30}
R(x\to 0)\to C_1\times x^{-{1\over 2}+\beta}+C_2\times x^{-{1\over
2}-\beta}\  .
\end{equation}
The boundary condition (\ref{Eq18}), when applied to the
near-horizon behavior (\ref{Eq30}) of the radial eigenfunction,
implies that regular bound-state scalar configurations are
characterized by the relations
\begin{equation}\label{Eq31}
C_2=0 \ \ \ \ \text{and}\ \ \ \ \beta\geq{1\over 2}\  .
\end{equation}
Note that (\ref{Eq31}) is consistent with our previous conclusion
(\ref{Eq27}).
\newline
(2) The asymptotic $x\to\infty$ behavior of the radial eigenfunction
(\ref{Eq29}) at spatial infinity is given by \cite{Abram}
\begin{eqnarray}\label{Eq32}
R(x\to\infty)&\to& C_1\times(-2\epsilon)^{-{1\over
2}-\beta+\kappa}{{\Gamma(1+2\beta)}\over{\Gamma({1\over
2}+\beta+\kappa)}}x^{-1+\kappa}e^{-\epsilon x} \nonumber
\\&& + C_1\times(2\epsilon)^{-{1\over
2}-\beta-\kappa}{{\Gamma(1+2\beta)}\over{\Gamma({1\over
2}+\beta-\kappa)}}x^{-1-\kappa}e^{\epsilon x}\ .
\end{eqnarray}
The boundary condition (\ref{Eq17}), when applied to the asymptotic
behavior (\ref{Eq32}) of the radial eigenfunction, implies that the
coefficient of the exploding exponent $e^{\epsilon x}$ in
(\ref{Eq32}) should vanish. This yields the characteristic resonance
equation \cite{Noteabr}
\begin{equation}\label{Eq33}
{1\over 2}+\beta-\kappa=-n\ \ \ \ \text{with}\ \ \ n=0,1,2,...\  .
\end{equation}
for the stationary bound-state resonances of the composed
Kerr-Newman-charged-massive-scalar-field system.

Taking cognizance of Eqs. (\ref{Eq2}), (\ref{Eq12}) and
(\ref{Eq23}), one can write $\kappa$ in the form
\begin{equation}\label{Eq34}
\kappa={{\alpha}\over{\epsilon}}-\epsilon\  ,
\end{equation}
where \cite{Notealpp}
\begin{equation}\label{Eq35}
\alpha\equiv m^2\cdot{{(s^2+\gamma)(1-\gamma)}\over{(1+s^2)^2}}\
\end{equation}
and \cite{Notegms}
\begin{equation}\label{Eq36}
\gamma\equiv {{qQs}\over{m}}\  .
\end{equation}
Likewise, taking cognizance of Eqs. (\ref{Eq2}), (\ref{Eq12}),
(\ref{Eq14}), and (\ref{Eq24}), one can write $\beta$ in the eikonal
$m\gg1$ regime in the form
\begin{equation}\label{Eq37}
\beta=\sqrt{\beta^2_0+\epsilon^2+\sqrt{m^2+(s\epsilon)^2}}\  ,
\end{equation}
where
\begin{equation}\label{Eq38}
\beta^2_0\equiv
m^2\cdot{{1-3s^2-4\gamma(1-s^2)+\gamma^2(3-s^2)}\over{(1+s^2)^2}}\
.
\end{equation}
Substituting (\ref{Eq34}) and (\ref{Eq37}) into (\ref{Eq33}), one
finds that the characteristic resonance condition can be written as
a (rather complicated) equation for the dimensionless physical
parameter $\epsilon$:
\begin{equation}\label{Eq39}
{1\over2}+\sqrt{\beta^2_0+\epsilon^2+\sqrt{m^2+(s\epsilon)^2}}-{{\alpha}\over{\epsilon}}+\epsilon=-n\
.
\end{equation}
From (\ref{Eq39}) one deduces that $\epsilon=\epsilon(n)$ is a {\it
decreasing} function of the resonance parameter $n$ \cite{Notedcn}.

It proves useful to write the resonance equation (\ref{Eq39}) in the
form
\begin{equation}\label{Eq40}
(\beta^2_0+2\alpha)\epsilon^2-\alpha^2=(n+1/2)[2\epsilon^3+(n+1/2)\epsilon^2-2\alpha\epsilon]
-m\sqrt{1+(s\epsilon/m)^2}\epsilon^2\  .
\end{equation}
Note that in the eikonal regime,
\begin{equation}\label{Eq41}
m\gg n+1/2\  ,
\end{equation}
the l.h.s of (\ref{Eq40}) is of order $O(m^4)$, whereas the r.h.s of
(\ref{Eq40}) contains terms of order $O[(n+1/2)m^3]$ and of order
$O[(n+1/2)^2m^2]$ [see Eqs. (\ref{Eq35}), (\ref{Eq38}), and
(\ref{Eq45}) below]. Thus, in the eikonal regime (\ref{Eq41}) one
can use an iteration scheme in order to solve the characteristic
resonance equation (\ref{Eq40}). The zeroth-order resonance equation
is given by $(\beta^2_0+2\alpha)\bar\epsilon^2-\alpha^2=0$, which
yields the simple zeroth-order ($n$-independent) solution
\begin{equation}\label{Eq42}
\bar\epsilon={{\alpha}\over{\sqrt{\beta^2_0+2\alpha}}}\  .
\end{equation}
Substituting (\ref{Eq42}) into the r.h.s of (\ref{Eq40}), one
obtains the first-order resonance equation
$(\beta^2_0+2\alpha)\epsilon^2_n-\alpha^2=(n+1/2)[2\bar\epsilon^3+(n+1/2)\bar\epsilon^2-2\alpha\bar\epsilon]
-m\sqrt{1+(s\bar\epsilon/m)^2}\bar\epsilon^2$, whose ($n$-dependent)
solution is given by
\begin{equation}\label{Eq43}
\epsilon_n=\bar\epsilon\cdot(1+\delta_n)\  ,
\end{equation}
where
\begin{equation}\label{Eq44}
\delta_n=-{{\beta^2_0+\alpha}\over{(\beta^2_0+2\alpha)^{3/2}}}\cdot(n+1/2)
-{{\sqrt{m^2+(s\bar\epsilon)^2}}\over{2(\beta^2_0+2\alpha)}}+O\Big[\Big({{n+1/2}\over{m}}\Big)^2\Big]\ll1\
\end{equation}
is a small correction factor [see Eq. (\ref{Eq46}) below].

Substituting the relations (\ref{Eq35}) and (\ref{Eq38}) into
(\ref{Eq42}) and (\ref{Eq44}), one finds \cite{Notegs,Noteops}
\begin{equation}\label{Eq45}
\bar\epsilon=m\cdot{{(s^2+\gamma)}\over{(1+s^2)\sqrt{1-s^2}}}\
\end{equation}
and
\begin{equation}\label{Eq46}
\delta_n=-{{(1+s^2)[1-2s^2-3\gamma(1-s^2)+\gamma^2(2-s^2)]}\over{(1-\gamma)^3(1-s^2)^{3/2}}}
\cdot{{n+1/2}\over{m}}-{{(1+s^2)\sqrt{1+s^2-s^4+2s^4\gamma+s^2\gamma^2}}\over{2(1-\gamma)^2(1-s^2)^{3/2}}}
\cdot{{1}\over{m}} =O\Big({{n+1/2}\over{m}}\Big)\ll1\ .
\end{equation}

Taking cognizance of Eqs. (\ref{Eq2}), (\ref{Eq12}), (\ref{Eq43}),
and (\ref{Eq45}), one finally finds the discrete spectrum
$\{\mu_n(m,qQ,s)\}^{n=\infty}_{n=0}$ of eigen field-masses which
characterize the stationary bound-state charged scalar clouds of the
extremal charged rotating Kerr-Newman black-hole spacetime:
\begin{equation}\label{Eq47}
\mu_n=\bar\mu\cdot(1+s^2\delta_n)\ \ \ ; \ \ \ n=0,1,2,...\ (n\ll
m)\ ,
\end{equation}
where \cite{Notergs}
\begin{equation}\label{Eq48}
M\bar\mu(m,qQ,s)=m\cdot{{s^2+\gamma}\over{s(1+s^2)\sqrt{1-s^2}}}\  .
\end{equation}

\section{Effective lengths (radii) of the stationary bound-state charged scalar clouds.}

In the present section we shall challenge the validity of the
no-short-hair bound (\ref{Eq1}) beyond the regime of
spherically-symmetric static black-hole spacetimes \cite{Noteprg}.
To this end, we shall now evaluate the effective lengths (radii) of
the {\it non}-spherically symmetric {\it non}-static charged scalar
clouds which characterize the extremal charged rotating Kerr-Newman
black-hole spacetimes.

Taking cognizance of Eqs. (\ref{Eq31}) and (\ref{Eq33}), one can
express the radial eigenfunctions (\ref{Eq29}) which characterize
the stationary bound-state charged scalar clouds in the form
\begin{equation}\label{Eq49}
R^{(n)}(x)=Ax^{\beta-{1\over 2}}e^{-\epsilon
x}L^{(2\beta)}_n(2\epsilon x)\  ,
\end{equation}
where $L^{(2\beta)}_n(x)$ are the generalized Laguerre polynomials
\cite{Notesea} and $A$ is a normalization constant.
In particular, the ground-state ($n=0$) radial eigenfunction is
given by the compact expression \cite{Noten0}
\begin{equation}\label{Eq50}
R^{(0)}(x)=Ax^{\beta-{1\over 2}}e^{-\epsilon x}\  .
\end{equation}

The radial eigenfunction (\ref{Eq50}), which characterizes the
fundamental (ground-state) charged scalar cloud, has a global
maximum at
\begin{equation}\label{Eq51}
x_{\text{peak}}={{\beta-{1\over2}}\over{\epsilon}}\  .
\end{equation}
Note that the expression (\ref{Eq51}) for the location of the peak
of the ground-state ($n=0$) radial eigenfunction (\ref{Eq50}) agrees
with the previously found expression (\ref{Eq28}) \cite{Notesub} for
the location of the minimum of the effective binding potential
(\ref{Eq22}).

Using the relation (\ref{Eq37}), one can write (\ref{Eq51}) in the
form
\begin{equation}\label{Eq52}
x_{\text{peak}}=\sqrt{1+({{\beta_0}/{\bar\epsilon}})^2}+O(m^{-1})\ .
\end{equation}
Substituting (\ref{Eq38}) and (\ref{Eq45}) into (\ref{Eq52}), one
finds
\begin{equation}\label{Eq53}
x_{\text{peak}}(\gamma;s)={{1-2s^2-\gamma(2-s^2)}\over{s^2+\gamma}}[1+O(m^{-1})]\
.
\end{equation}
This expression for the peak location of the radial eigenfunction
(\ref{Eq50}) provides a quantitative measure for the characteristic
size (length) of the fundamental bound-state charged scalar cloud.

At this point, it is interesting to note that {\it neutral} scalar
clouds, which are characterized by the simple relation
\cite{Hodaa,Notebts}
\begin{equation}\label{Eq54}
x_{\text{peak}}(\gamma=0;s)={{1-2s^2}\over{s^2}}[1+O(m^{-1})]\ ,
\end{equation}
respect the no-short-hair lower bound (\ref{Eq1}). To see this, we
recall that the radii of the equatorial ($l=m\gg1$) null circular
geodesics which characterize the charged rotating Kerr-Newman
black-hole spacetimes are given by the relation \cite{Chan}
\begin{equation}\label{Eq55}
r^2-3Mr+2Q^2+2a(Mr-Q^2)^{1/2}=0\  .
\end{equation}
This equation can be solved analytically for near-extremal black
holes. In particular, one finds
\begin{equation}\label{Eq56}
r_{\text{null}}=
\begin{cases}
2(M-a)+O(\sqrt{M^2-a^2-Q^2}) & \text{for}\ \ \ 0\leq a/M\leq 1/2 \ \
\ ;
\\ M+O(\sqrt{M^2-a^2-Q^2}) & \text{for}\ \ \ 1/2<
a/M\leq 1\  ,
\end{cases}
\end{equation}
which implies [see Eq. (\ref{Eq21})]
\begin{equation}\label{Eq57}
x_{\text{null}}(s)\to
\begin{cases}
1-2s & \text{for}\ \ \ 0\leq s\leq 1/2 \ \ \ ;
\\ 0 & \text{for}\ \ \ 1/2<
s\leq 1\
\end{cases}
\end{equation}
in the extremal $M^2-a^2-Q^2\to 0$ limit.

Taking cognizance of Eqs. (\ref{Eq54}) and (\ref{Eq57}), one finds
the characteristic inequality
\begin{equation}\label{Eq58}
x_{\text{peak}}(\gamma=0;s)>x_{\text{null}}(s)
\end{equation}
for neutral scalar clouds in the entire range $s\in
(0,{1\over\sqrt{2}})$
\cite{Notebts}. We therefore conclude that {\it neutral} scalar
clouds conform to the no-short-hair lower bound (\ref{Eq1}). In
particular, for the neutral scalar clouds one finds
$\text{min}_s\{{{x_{\text{peak}}(\gamma=0;s)}/{x_{\text{null}}(s)}}\}=10+6\sqrt{3}\simeq20.392$
for $s=(\sqrt{3}-1)/2\simeq0.366$.

Let us now return to the more general case of {\it charged} scalar
clouds. From the expression (\ref{Eq53}) for
$x_{\text{peak}}(\gamma;s)$ one deduces that, for a given value of
the black-hole angular momentum $s$, $x_{\text{peak}}(\gamma;s)$ is
a {\it decreasing} function of the dimensionless physical parameter
$\gamma$. In particular, the expression (\ref{Eq53}) for
$x_{\text{peak}}(\gamma;s)$ reveals the remarkable fact that the
exterior charged scalar clouds can be made arbitrary compact. In
particular, one finds \cite{Notegc}
\begin{equation}\label{Eq59}
x_{\text{peak}}(s)\to 0 \ \ \ \ \text{for}\ \ \ \
\gamma(s)\to\gamma^*(s)={{1-2s^2}\over{2-s^2}}\  .
\end{equation}
It is worth emphasizing that the relation (\ref{Eq59}) is valid for
extremal Kerr-Newman black holes in the {\it entire} range $s\in
(0,1)$ of the black-hole rotation parameter. This fact implies that
{\it all} \cite{Notead} charged and rotating extremal Kerr-Newman
black holes can support extremely compact charged scalar
configurations in their exterior regions \cite{Noteqo}.

Taking cognizance of Eqs. (\ref{Eq57}) and (\ref{Eq59}), one finds
the important inequality
\begin{equation}\label{Eq60}
x_{\text{peak}}(\gamma=\gamma^*;s)<x_{\text{null}}(s)
\end{equation}
for charged scalar clouds in the entire range $s\in (0,{1\over2})$.
We therefore conclude that {\it charged} scalar clouds may violate
the no-short-hair lower bound (\ref{Eq1}) \cite{Noteprg,Notefr}.

\section{Summary.}

A no-short-hair theorem for spherically-symmetric static black holes
was proved in \cite{Hod11}. This theorem reveals that, if a {\it
spherically}-symmetric {\it static} black hole has hair, then this
hair (i.e. the external fields) must extend beyond the null circular
geodesic (the``photonsphere") of the corresponding black-hole
spacetime [see Eq. (\ref{Eq1})].

The main goal of the present study was to challenge the validity of
this no-short-hair property beyond the regime of static
spherically-symmetric hairy black-hole configurations. To that end,
we have studied analytically the physical properties of extremal
charged rotating Kerr-Newman black holes linearly coupled to
non-spherically symmetric stationary (rather than static) charged
massive scalar fields.

In particular, for given parameters $\{M,Q,J\}$ of the central
Kerr-Newman black hole, we have identified the critical value of the
field charge coupling constant, $q=q^*(s)$ [see Eqs. (\ref{Eq36})
and (\ref{Eq59})], which minimizes the effective radial lengths of
the exterior stationary charged scalar configurations. This allowed
us to prove that the ({\it non}-static, {\it non}-spherically
symmetric) composed Kerr-Newman-charged-massive-scalar-field
configurations can violate the no-short-hair lower bound
(\ref{Eq1}). In particular, it was shown that extremal Kerr-Newman
black holes in the {\it entire} parameter space $s\in (0,1)$ can
support extremely compact stationary charged scalar clouds (made of
linearized charged massive scalar fields with the property
$r_{\text{field}}\to r_{\text{H}}$) in their exterior regions.
Specifically, these short-range
Kerr-Newman-charged-massive-scalar-field configurations are
characterized by the simple relation [see Eqs. (\ref{Eq48}) and
(\ref{Eq59})]
\begin{equation}\label{Eq61}
x_{\text{peak}}(s)\to 0 \ \ \ \ \text{for}\ \ \ \
{{q}\over{\mu}}\to{{1-2s^2}\over{1-s^2}}\  .
\end{equation}
It is interesting to note that, in order to support these extremely
compact field configurations, the extremal black hole must have {\it
both} angular-momentum {\it and} electric charge \cite{Notesoi}.

Finally, it is worth emphasizing again that we have treated here the
exterior charged massive scalar fields at the linear level. As we
have shown, the main advantage of this approach lies in the fact
that the composed Kerr-Newman-linearized-charged-scalar-field system
is amenable to an {\it analytical} treatment. We believe that it
would be highly important to use {\it numerical} techniques
\cite{HerR} in order to generalize our results to the regime of
non-linear exterior fields.

\bigskip
\noindent
{\bf ACKNOWLEDGMENTS}
\bigskip

This research is supported by the Carmel Science Foundation. I thank
Yael Oren, Arbel M. Ongo, Ayelet B. Lata, and Alona B. Tea for
stimulating discussions.



\begin{thebibliography}{99}
\bibitem{Noteflat} We refer here to asymptotically flat black holes.
\bibitem{Notevio} It is worth noting that most of these
hairy black-hole solutions are known to be dynamically unstable.
\bibitem{Noterc} It is worth emphasizing that, for the superradiant amplification to occur,
the black-hole spacetime must possess angular momentum and/or
electric charge.
\bibitem{Noteunits} We use natural units in which $G=c=\hbar=1$.
\bibitem{Notepar} Here $a, Q$, and $r_+$ are respectively the angular momentum per unit mass,
electric charge, and horizon-radius of the charge rotating
Kerr-Newman black-hole spacetime.
\bibitem{Notesmp} One can assume $s>0, Q>0$, and $m>0$ without loss of
generality.
\bibitem{Notedim} Note that the mass and charge parameters of the scalar field, $\mu$ and $q$,
stand respectively for $\mu/\hbar$ and $q/\hbar$. Hence, these
physical parameters have the dimensions of $($length$)^{-1}$.
\bibitem{Notemm} Here $M$ is the mass of the central Kerr-Newman black hole.
\bibitem{Noteanz} Here $\omega,\ m$, and $l\geq|m|$ are respectively the conserved
frequency and the harmonic indices (azimuthal and spheroidal) of the
field mode [see Eq. (\ref{Eq10}) below].
\bibitem{Noteangu} Note that the spheroidal harmonic eigenfunctions $S_{lm}(\theta;{s}\epsilon)$
reduce to the familiar spherical harmonic functions in the
$s\epsilon\ll 1$ limit. In this case one finds the familiar angular
eigenvalues $K_{mm}(s\epsilon=0)=m(m+1)$.
\bibitem{Notebr} We shall henceforth omit the angular harmonic indices $\{l,m\}$ for brevity.
\bibitem{Noteepp} One may choose $\epsilon>0$ without loss of generality.
\bibitem{Notenr} Here $n=0,1,2,...$ is the resonance parameter, see
Eq. (\ref{Eq33}) below.
\bibitem{Note1o2} The limiting case of $\beta^2=1/4$ with $V(x\to 0)\sim -1/x$ is also allowed.
In this special case the effective binding potential (\ref{Eq22})
takes the form of a radial Coulomb potential with no centrifugal
barrier. Thus, in this special case one finds that the radial
function $\psi(x)$ is described by the familiar
spherically-symmetric ($l=0$) eigenfunctions of the hydrogen atom.
In particular, as for the spherically-symmetric eigenfunctions of
the hydrogen atom, one finds $\psi(x\to 0)\sim x^1$ [or
equivalently, $R(x\to 0)\to\text{constant}$, see Eqs. (\ref{Eq30})
and (\ref{Eq31}) below] in this special case.
\bibitem{Notebeta} One may choose $\beta>0$ without loss of generality.
\bibitem{Noteabr} Here we have used the well-known pole structure of the Gamma
functions, see equation 6.1.7 of \cite{Abram}.
\bibitem{Notealpp} Taking cognizance of Eqs. (\ref{Eq31}) and
(\ref{Eq33}), one finds the necessary requirement
$\kappa=(n+1/2)+\beta>0$ for the existence of the  stationary
bound-state charged scalar configurations. This implies that
$\alpha>\epsilon^2>0$ [see Eq. (\ref{Eq34})].
\bibitem{Notegms} Taking cognizance of Eq. (\ref{Eq2}), one finds
that the requirement $\omega_{\text{field}}=\omega_{\text{c}}>0$
\cite{Zel,PressTeu1} implies $ms+qQ>0$, or equivalently
$\gamma>-s^2$.
\bibitem{Notedcn} From the relation
$n=-{1\over2}-\sqrt{\beta^2_0+\epsilon^2+\sqrt{m^2+(s\epsilon)^2}}+{{\alpha}/{\epsilon}}-\epsilon$
one finds
$dn/d\epsilon=-\{[1+s^2/2\sqrt{m^2+(s\epsilon)^2}]\epsilon/\beta+{{\alpha}/{\epsilon}^2}+1\}$
which is negative definite (see Eq. (\ref{Eq31}) and
\cite{Notealpp}). Thus, $d\epsilon/dn<0$, which implies that
$\epsilon=\epsilon(n)$ is a decreasing function of the resonance
parameter $n$.
\bibitem{Notegs} Note that the requirement $\alpha-\epsilon^2>0$ (see
\cite{Notealpp}) implies the inequality
$\gamma<(1-2s^2)/(2-s^2)+O(m^{-1})$ for the charged scalar clouds
[see Eqs. (\ref{Eq35}) and (\ref{Eq45})].
\bibitem{Noteops} In the opposite regime $n+1/2\gg m$ of excited eigenstates, one finds from
(\ref{Eq39}) the simple asymptotic solution
$\epsilon_n=\alpha/n\ll1$, which implies
$\mu_n=\omega_{\text{c}}\{1+[ms(1-\gamma)/(1+s^2)n]^2/2\}$ [see Eq.
(\ref{Eq12})].
\bibitem{Notergs} Note that the resonance equation (\ref{Eq39}) forbids $\epsilon=0$,
which implies $s\neq0$ [The requirement $\epsilon\neq0$ stems from
the physical boundary condition (\ref{Eq17}). In particular, from
(\ref{Eq17}) one learns that an exponentially decaying radial
eigenfunction at asymptotic infinity requires $\epsilon>0$].
Likewise, the resonance equation (\ref{Eq39}) forbids
$\epsilon=\infty$, which implies $s\neq1$ (The inequality
$\epsilon<\infty$ reflects the fact that the allowed scalar field
masses $\mu$ are finite).
\bibitem{Notesea} Here we have used equation $13.6.9$ of \cite{Abram}.
\bibitem{Noten0} Here we have used the fact that $L^{(2\beta)}_0(x)$ is a constant independent of $x$.
\bibitem{Notebts} Note that the requirement $\alpha-\epsilon^2>0$ (see
\cite{Notealpp}) implies the inequality
$s<{{1}\over{\sqrt{2}}}+O(m^{-1})$ for the neutral scalar clouds
[see Eqs. (\ref{Eq35}) and (\ref{Eq45})].
\bibitem{Notegc} This value of $\gamma^*(s)$ corresponds to
$Mq(s)\to Mq^*(s)=m\cdot{{1-2s^2}\over{s(2-s^2)\sqrt{1-s^2}}}$.
Taking cognizance of Eqs. (\ref{Eq48}) and (\ref{Eq59}), one finds
the dimensionless ratio $Qq^*(s)/M\bar\mu(s)=(1-2s^2)/\sqrt{1-s^2}$.
Note that $Qq^*(s)/M\bar\mu(s)<1$ in the entire range $s\in (0,1)$.
\bibitem{Notefr} It is important to emphasize the fact that the
results of \cite{Hodrc,HerR} indicate that scalar clouds with
reasonably small values of the angular indices $\{l,m\}$ and with
reasonably small values of the dimensionless quantity $qQ$ respect
the no-short hair conjecture (\ref{Eq1}).
\bibitem{Notesoi} Thus, the values $s=0$ and $s=1$ for the extremal black-hole angular momentum are excluded
\cite{Notergs}.
\end{thebibliography}
\end{document}